\documentclass[prl,aps,graphics,epsfig,twocolumn,floats,superscriptaddress,showpacs]{revtex4}
\hfuzz=\maxdimen
\usepackage{graphicx}
\usepackage{amssymb}
\usepackage{amsmath}
\usepackage{color}
\usepackage{mathptmx}

\begin{document}

\title{Nature of magnetotransport in metal/insulating-ferromagnet heterostructures: Spin Hall magnetoresistance or magnetic proximity effect}

 \author{X. Zhou}
\affiliation{Shanghai Key Laboratory of Special Artificial Microstructure Materials and Technology and Pohl Institute of Solid State Physics and School of Physics Science and Engineering, Tongji University, Shanghai 200092, China}
\author{L. Ma}
\affiliation{Shanghai Key Laboratory of Special Artificial Microstructure Materials and Technology and Pohl Institute of Solid State Physics and School of Physics Science and Engineering, Tongji University, Shanghai 200092, China}
\author{Z. Shi}
\affiliation{Shanghai Key Laboratory of Special Artificial Microstructure Materials and Technology and Pohl Institute of Solid State
Physics and School of Physics Science and Engineering, Tongji University, Shanghai 200092, China}
\author{W. J. Fan}
\affiliation{Shanghai Key Laboratory of Special Artificial Microstructure Materials and Technology and Pohl Institute of Solid State
Physics and School of Physics Science and Engineering, Tongji University, Shanghai 200092, China}
\author{Jian-Guo Zheng}
\affiliation{Irvine Materials Research Institute, University of California, Irvine, CA 92697-2800, USA}
\author{R. F. L. Evans}
\affiliation{Department of Physics, University of York, York YO10 5DD, United Kingdom}
 \author{S. M. Zhou}
\affiliation{Shanghai Key Laboratory of Special Artificial Microstructure Materials and Technology and Pohl Institute of Solid State Physics and School of Physics Science and Engineering, Tongji University, Shanghai 200092, China}

\date{\today}
\vspace{5cm}

\begin{abstract}
We study the anomalous Hall-like effect (AHLE) and the effective anisotropic magnetoresistance (EAMR) in antiferromagnetic $\gamma$-IrMn$_{3}$/Y$_{3}$Fe$_{5}$O$_{12}$(YIG) and Pt/YIG heterostructures. For $\gamma$-IrMn$_{3}$/YIG, the EAMR and the AHLE resistivity change sign with temperature due to the competition between the spin Hall magnetoresistance (SMR) and the magnetic proximity effect (MPE) induced by the interfacial antiferromagnetic uncompensated magnetic moment. In contrast, for Pt/YIG the AHLE resistivity changes sign with temperature whereas no sign change is observed in the EAMR. This is because the MPE and the SMR play a dominant role in the AHLE and the EAMR, respectively. As new types of galvanomagnetic property, the AHLE and the EAMR  have proved vital in disentangling the MPE and the SMR in metal/insulating-ferromagnet heterostructures.
\end{abstract}


\maketitle

\indent Since the first observation of spin Hall effect (SHE) in semiconductors, it has been studied extensively because of intriguing physics and important applications in generation and detection of pure spin currents~\cite{Kato2004,Wunderlich2005,Guo2008,Hoffmann2014}. The SHE in heavy nonmagnetic metal (NM) strongly depends on the electronic band structure and the spin orbit coupling (SOC)~\cite{Guo2008}. The inverse spin Hall effect (ISHE) enables to electrically detect the spin current~\cite{Valenzuela2006}. In the spin pumping technique, for example, the ISHE is employed to detect the spin current in a NM layer by measuring the transverse voltage when the magnetization precession of a neighboring ferromagnet (FM) layer is excited~\cite{Ando2011,Czeschka2011}. \\ 
\indent In their pioneering work, Nakayama~\textit{et al.} proposed spin Hall magnetoresistance (SMR) in NM/insulating-FM as a way to study the SHE in heavy NM~\cite{Nakayama2013}. Since then, the SMR has attracted a lot of attention~\cite{Althammer2013,Vlietstra2013}. When a charge current is applied in the NM layer, a spin current is produced along the film normal direction due to the SHE and the reflected spin current is modified by the orientation of the underlying FM magnetization with respect to the charge current. Since the reflected spin current produces an additional electric field through the ISHE, the measured resistivity of the NM layer strongly depends on the orientation of the FM magnetization. The longitudinal and the transverse resistivity read~\cite{Nakayama2013}:
\begin{equation}
\rho_{xx}=\rho_0 +\rho_{1}m^2_{t},~\rho_{xy}=-\rho_{1}m_tm_j+\rho_{2}m_n,
\label{pauli1}
\hspace{0.0 cm}
\end{equation}
 where $m_n$ is the component of the magnetization unit vector along the film normal direction, and the in-plane components $m_j$ and $m_t$ are parallel to and perpendicular to the sensing charge current, respectively. Being negative, parameters $\rho_1$ and $\rho_2$ refer to the spin Hall induced anisotropic magnetoresistance (SH AMR) and anomalous Hall effect (SH AHE), respectively. However, Huang \textit{et al} found that the magnetic proximity effect (MPE) may be involved~\cite{Huang2011,Huang2012,QU2013,Lu2013,JShi2014,Miao2014}. For the spin polarized NM layer,
 the magnetoresistance (MR) effect occurs as observed in conventional metallic FMs~\cite{Mcguire1975}:
\begin{equation}
\rho_{xx}=\rho_0+\Delta\rho_{AMR}m^2_{j},~\rho_{xy}=\Delta\rho_{AMR}m_tm_j+\rho_{AHE}m_n,
\label{pauli2}
\hspace{0.0 cm}
\end{equation}
 where $\rho_{AHE}$ and $\Delta\rho_{AMR}$ correspond to the MPE induced anomalous Hall effect (MPE AHE) and the MPE AMR, respectively. The emerging MPE makes it complicated to clarify the mechanism of either the MR phenomena in NM/insulating-FM or the SHE in the NM layer~\cite{Nakayama2013,Althammer2013,Vlietstra2013,Huang2011,Huang2012,QU2013,Lu2013}. \\
\indent  With the external magnetic field $H$ along the film normal direction, the Hall resistivity in the NM layer exhibits a similar magnetic field dependence for the AHE in bulk metallic FMs, exhibiting the anomalous Hall-like effect (AHLE).
Since the two components, the MPE AHE and the SH AHE, are of an \textit{interfacial} nature, unlike the bulk feature of the conventional AHE, the AHLE is expected to bring new interesting information. For Pt/Y$_{3}$Fe$_{5}$O$_{12}$(YIG), for example, the AHLE resistivity $\rho_{AHLE}$ changes sign with temperature $T$~\cite{Huang2012,Shimizu2013,XZhou2014} whereas
for Pd/YIG it is positive for $T$ in the region from 5 K to 300 K~\cite{XZhou2014}. In a similar way, the effective AMR (EAMR) can be defined when $H$ is rotated in the $xy$ plane.\\
\indent In this Letter, we study AHLE and EAMR in $\gamma$-IrMn$_{3}$(=IrMn)/YIG and Pt/YIG to determine the role of the MPE, where IrMn and Pt layers are antiferromagnetic and nearly ferromagnetic, respectively, exhibiting different magnetic attributes. With a strong SOC of heavy Ir atoms and a low magnetic damping parameter in the insulating YIG layer, a sizable SMR effect is expected in IrMn/YIG. Meanwhile, exchange bias(EB) can be established below the blocking temperature $T_B$ and a small uncompensated magnetic moment may be induced~\cite{Zhang2014}, exhibiting an effect similar to the MPE. Accordingly, the MPE occurs at low $T$ and disappears at high $T$. For Pt/YIG, however, the MPE exists at all $T$. The \textit{different} $T$ dependencies of the MPE in the two hybrid structures allow the SMR and the MPE to be separated and demonstrate the important role of AHLE and EAMR in studies of the intricate MR properties in NM/insulating-FM heterostructures.\\
\begin{figure}[tb]
\begin{center}
\includegraphics[width=5cm]{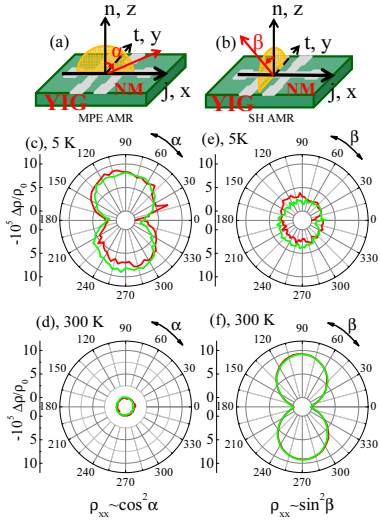}
\caption{Measurement geometries of the MPE AMR (a) and the SH AMR (b). In (a, b), the sensing electric current is applied along the \emph{x} axis. Angular dependent $\Delta\rho/\rho_0$ in the \emph{xz} (c, d) and \emph{yz}(e, f) planes at 5 K (c, e) and 300 K (d, f) for IrMn/YIG. Here, the red and green lines refer to the clockwise and counter clockwise rotations of the external magnetic field $H=10$ kOe, and $\Delta\rho=\rho_{xx}-\rho_0$.} \label{Fig1}
\end{center}
\end{figure}
\begin{figure}[tb]
\begin{center}
\includegraphics[width=5cm]{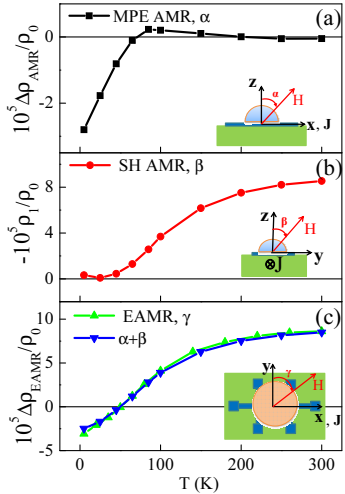}
\caption{For IrMn/YIG, the $T$ dependencies of $\Delta\rho_{AMR}/\rho_0$ (a), $-\rho_1/\rho_0$ (b), and $\Delta\rho_{EAMR}/\rho_0$ (c). The data in (a, b, c) were achieved from measurements of angular dependence in \emph{xz}, \emph{yz}, and \emph{xy} planes under $H=10$ kOe, respectively. In (c), the sum of $-\rho_1/\rho_0$ and $\Delta\rho_{AMR}/\rho_0$ is also given.} \label{Fig2}
\end{center}
\end{figure}
\indent  IrMn (2.5 nm)/YIG (20 nm) and Pt (2.5 nm)/YIG (20 nm) heterostructures were fabricated by pulsed laser deposition and subsequent DC magnetron sputtering in an ultrahigh vacuum system on (111)-oriented, single crystalline Gd$_3$Ga$_5$O$_{12}$ (GGG) substrates. Epitaxial growth of the YIG layer was proved by x-ray diffraction at high angles and pole figure, and by transmission electronic microscopy (TEM), as shown in Figs.S1 and S2 in supplementary materials~\cite{Supple}. The thickness of the YIG layer was characterized by x-ray reflection (XRR). Magnetization hysteresis loops of the samples were measured using the physical property measurement system (PPMS). After the films were patterned into a normal Hall bar, the longitudinal resistivity ($\rho_{xx}$) and the transverse Hall resistivity ($\rho_{xy}$) were measured by PPMS. \\
\indent When the magnetic field $H$ is applied in the \emph{xz} plane in Fig.~\ref{Fig1}(a), $\rho_{xx}$ approximately shows the $\cos^2\alpha$ angular dependence at low $T$, i.e., $\rho_{xx}\simeq\rho_0+\Delta\rho_{AMR}\cos^2\alpha$, but it has no variation at high $T$, as shown in Figs.~\ref{Fig1}(c)~and~\ref{Fig1}(d). The oscillation amplitude $\Delta\rho_{AMR}$ decreases with increasing $T$ and vanishes at high $T$. With $m_t\equiv0$ in the \emph{xz} plane, the SH AMR is excluded and above results arise from the MPE AMR which is in turn accompanied by EB at low $T$, as shown in Fig.S3 in the supplementary materials~\cite{Supple}. Atomistic simulations\cite{Vampire} confirm the presence of an uncompensated magnetic moment at the IrMn/FM interface, shown in Fig.~S4 in the supplementary materials~\cite{Supple}. Due to the structural degradation induced by the lattice mismatch between IrMn and YIG layers, $T_B$ of 100 K in the ultrathin IrMn layer is much lower than the N\'{e}el temperature (400-520 K) of bulk IrMn~\cite{Nogues1999}. When $H$ is rotated in the \emph{yz} plane in Fig.~\ref{Fig1}(b), $m_j\equiv0$, the MPE is excluded, and the results in Figs.~\ref{Fig1}(e) and ~\ref{Fig1}(f) correspond to the SH AMR. At high $T$, $\rho_{xx}$ has the $\sin^2\beta$ angular dependence, i.e, $\rho_{xx}=\rho_0+\rho_1\sin^2\beta$, whereas $\rho_{xx}$ has no variation at low $T$, that is to say, the oscillatory amplitude $|\rho_1|$ increases with increasing $T$.\\
\indent Figure~\ref{Fig2}(a) shows that the ratio $\Delta\rho_{AMR}/\rho_0$ increases from negative to positive and finally approaches zero as $T$ changes from 5 K to 300 K. This phenomenon stems from the measurement strategy in which $\Delta\rho_{AMR}$ is obtained by the angular dependence of $\rho_{xx}$ and contributed by three different mechanisms. Induced by the uncompensated magnetic moment, the first effect, MPE AMR, appears at low $T$ and vanishes at $T>T_B$. The second effect is caused by the forced magnetization induced MR under high $H$. The uncompensated magnetic moment at finite $T$ favors alignment under high $H$, leading to a negative MR. Near $T_B$, the second one becomes prominent and then vanishes at $T>T_B$. Caused by the ordinary MR, the third term is always positive for all $T$ and becomes weak when the mean free path becomes short at high $T$. Figure~\ref{Fig2}(b) shows that the ratio $-\rho_1/\rho_0$ becomes large in magnitude at high $T$. Apparently, the SH AMR and the MPE AMR become strong and weak with increasing $T$, respectively. Interestingly, Fig.~\ref{Fig2}(c) shows that $\Delta\rho_{EAMR}/\rho_0$ measured in the \emph{xy} plane in which $m_n\equiv0$ and $m_j^2+m_t^2\equiv1$, is approximately equal to the sum of $\Delta\rho_{AMR}/\rho_0$ and $-\rho_1/\rho_0$. As observed in Pd/YIG~\cite{JShi2014}, one has the following equation according to Eqs.~\ref{pauli1} and~\ref{pauli2},
\begin{equation}
\Delta\rho_{EAMR}=\Delta\rho_{AMR}-\rho_1,
\label{pauli3}
\hspace{5.0 cm}
\end{equation}
In particular, $\Delta\rho_{EAMR}$ also changes from negative to positive as a function of $T$, indicating the competition between the MPE and the SMR. \\
\indent Figure~\ref{Fig3}(a) shows that the angular dependencies of $\rho_{xy}$ in the \emph{xz} and \emph{yz} planes are identical, in agreement with Eqs.~\ref{pauli1} and~\ref{pauli2}. Since the value of $\rho_{AHLE}$ includes the ordinary Hall effect (OHE)~\cite{Althammer2013}, and the OHE at $H=10$ kOe might be much larger than the value of $\rho_2$, it is necessary to exclude the OHE contribution. In order to rigorously achieve $\rho_{AHLE}$, the Hall loops were measured for all samples, as shown in Fig.~\ref{Fig3}(b). Here, $\rho_{AHLE}=(\rho_{xy+}-\rho_{xy-})/2$, where $\rho_{xy+}$ and $\rho_{xy-}$ are extrapolated from positive and negative saturations, respectively. Significantly, the AHLE angle $\rho_{AHLE}/\rho_0$ also changes sign near $T=100$ K, as shown in Fig.~\ref{Fig3}(c). Since $\rho_2$ is always negative~\cite{Nakayama2013}, the sign change cannot be explained only in terms of the SH AHE, and the MPE AHE should also be considered according to the following equation:
\begin{equation}
\rho_{AHLE}=\rho_{AHE}+\rho_2.
\label{pauli4}
\hspace{5.0 cm}
\end{equation}
\indent It is revealing to analyze the sign changes of $\rho_{AHLE}$ and $\Delta\rho_{EAMR}$ in IrMn/YIG. The EB is established at $T<T_B$, as shown in supplementary materials~\cite{Supple}. It is also evidenced by the rotational hysteresis loss between clockwise and counter clockwise curves in Figs.~\ref{Fig1}(c),~\ref{Fig1}(e), and~\ref{Fig3}(a). Accordingly, the MPE AHE and the MPE AMR are induced by the uncompensated magnetic moment at $T<T_B$~\cite{Zhang2014}, i.e., $\rho_{AHE}\neq0$ and $\Delta\rho_{AMR}\neq0$, and they disappear at $T>T_B$. Meanwhile, the SH AHE and the SH AMR, i.e., $\rho_2$ and $\rho_1$, are small at low $T$ and become large  in magnitude at high $T$. Apparently, both $\rho_{AHLE}$ and $\Delta\rho_{EAMR}$ are mainly contributed by the MPE at low $T$ and the SMR at high $T$, respectively. Since the signs of $\Delta\rho_{AMR}$ and $\rho_{AHE}$ are opposite to those of $-\rho_1$ and $\rho_2$, the sign changes of $\rho_{AHLE}$ and $\Delta\rho_{EAMR}$ can therefore be easily understood. \\
\indent Without the data of the spin diffusion length, it is difficult to separate $\rho_2$ and $\rho_{AHE}$ in IrMn/YIG. At 5 K, however, $\rho_2$ is expected to be zero due to vanishing $\rho_1$ in Fig.~\ref{Fig2}(b), and $\rho_{AHE}$ is approximately equal to the measured $\rho_{AHLE}$, i.e., $\rho_{AHE}\simeq\rho_{AHLE}=2.0\times10^{-3}$ $\mu\Omega\mathrm{cm}$. The anomalous Hall conductivity (AHC) in the ultrathin IrMn layer is $\sigma_{AHC}=-0.045$ S/cm, much smaller than the calculated results of bulk IrMn (200-400 S/cm) based on the model of nonlinear antiferromagnetism~\cite{21Chen2014}. Since $\rho_{AHE}$ at 5 K decreases sharply with the IrMn layer thickness, as shown in Fig.S5~\cite{Supple}, the MPE AHE at low $T$ is proved to originate from the interfacial IrMn uncompensated magnetic moment and other physical sources can be excluded. Furthermore, near $T=300$ K, the MPE AHE disappears, i.e., $\rho_{AHE}=0$, and $\rho_2$ is thus equal to the measured $\rho_{AHLE}$, i.e., $\rho_2=\rho_{AHLE}=1.76\times10^{-3}$ $\mu\Omega\mathrm{cm}$. \\
\begin{figure}[tb]
\begin{center}
\includegraphics[width=5cm]{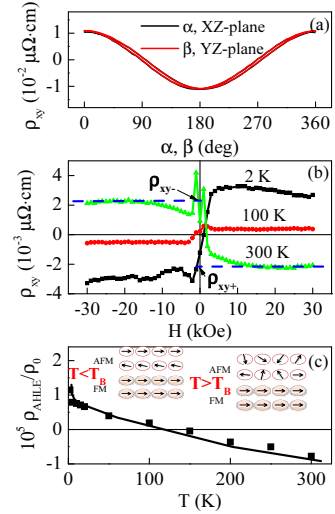}
\caption{For IrMn/YIG, the angular dependence of $\rho_{xy}$ in the \emph{xz} (black line) and \emph{yz} (red line) planes at 5 K and $H=10$ kOe (a), the Hall loops under $H$ along the film normal direction at $T=2$ K, 100 K, and 300 K (b), and the AHLE angle $\rho_{AHLE}/\rho_0$ versus $T$ (c). The left and right insets in (c) schematically show the spin structure in the IrMn layer below and above $T_B$, respectively.} \label{Fig3}
\end{center}
\end{figure}
\indent Figure~\ref{Fig4}(a) shows $\rho_{AHLE}$ and $\rho_2$ in Pt(2.5 nm)/YIG. Here, $\rho_2$ was calculated in the frame of the SMR theory~\cite{Nakayama2013}, with the film thickness (2.5 nm) of Pt, the ratio of real and imaginary parts of the spin mixing conductance at Pt/YIG interface~\cite{Althammer2013}, i.e., $G_i/G_r=0.03$, the spin diffusion length in the inset of Fig.~\ref{Fig4}(a)~\cite{Marmion2014}, and the measured $\rho_1$ in Fig.~\ref{Fig4}(b). Since $|\rho_2|\ll|\rho_{AHLE}|$ at \textit{all} $T$, the sign change of $\rho_{AHLE}$ cannot be explained in terms of the SH AHE, and instead it is mainly caused by the MPE AHE according to Eq.~\ref{pauli4}. It is noted that no sign change was observed in $\rho_{AHLE}$ for nearly-ferromagnetic-Pd/YIG~\cite{XZhou2014}. The AHLE behavior in Pt/YIG is different from those of Pd/YIG~\cite{JShi2014,XZhou2014} and IrMn/YIG. The $T$ dependence of $\rho_{AHLE}$ in NM/insulating-FM hybrid structure relies not only on the magnetic attribute in the NM layer but also on the electronic band structures near the Fermi level~\cite{Guo2008,Guo2014}. Shimizu \textit{et al.}, for example, found that the $T$ dependence of the AHLE in Pt/YIG can be tuned by the gate voltage~\cite{Shimizu2013}. With \textit{ab. initio} calculation results~\cite{Guo2014,Guo2008}, the magnetic moment of Pt atoms is evaluated to be as small as 0.003 $\mu_B$ with $\sigma_{AHC}=2.0$ S/cm at 5 K. Although the magnetic moment of Pt atoms depends on the chemical state on the YIG surface and the orbital hybridization of Fe and Pt atoms, it is generally smaller than the resolution ($\sim0.01$ $\mu_B$) of x-ray magnetic circular dichroism and hard to be accurately detected with this technique~\cite{XMCDAPL,Lu2013}.\\
\indent Figure~\ref{Fig4}(b) shows for Pt/YIG, $|\Delta\rho_{AMR}|\ll|\Delta\rho_{EAMR}|$ and thus $|\Delta\rho_{EAMR}|\simeq|\rho_1|$ for \textit{all} $T$. Accordingly, Eq.~\ref{pauli3} also holds for this system~\cite{JShi2014}. Moreover, $-\rho_1$ and $\Delta\rho_{EAMR}$, both being positive, change non-monotonically with $T$ in Fig.~\ref{Fig4}(b), as observed in Pd/YIG and PtPd/YIG~\cite{XZhou2014,JShi2014}. This is because the SH AMR changes non-monotonically with the spin diffusion length which changes monotonically with $T$ as shown in the inset of Fig.~\ref{Fig4}(a)~\cite{Marmion2014,Chen2013}. Consequently, $-\rho_1$ and $\Delta\rho_{EAMR}$ were also found to change non-monotonically with the Pt layer thickness~\cite{Huang2012,QU2013,Althammer2013}. The results in Fig.~\ref{Fig4} unambiguously show the dominant role of the MPE (SMR) in the AHLE (EAMR) in Pt/YIG. Alternatively, the MPE (SMR) is negligibly small in the EAMR (AHLE). Therefore, we \textit{solve} the dispute over the mechanism of the mixed MR in this system~\cite{Nakayama2013,Althammer2013,Vlietstra2013,Huang2011,Huang2012,QU2013,Lu2013}. \\
\indent It is significant to compare the SMR in IrMn/YIG and Pt/YIG. At 300 K, the spin Hall angle is reported to be about 0.028 and 0.056 for IrMn and Pt, respectively~\cite{Mendes2014,Rojas2014}. Since the spin dependent scattering in Pt (IrMn) is induced by the strong SOC (both strong SOC and magnetic ordering), the magnitude of the spin diffusion length and its $T$ dependence may be different in IrMn and Pt. For example, it is 0.295 nm for IrMn and 0.5-3.4 nm for Pt at 300 K~\cite{Acharya2011,Rojas2014}. Accordingly, $\rho_1$ in IrMn/YIG and Pt/YIG exhibits different variation trends with $T$, as shown in Fig.~\ref{Fig2}(b) and Fig.~\ref{Fig4}(b). With measured $\rho_1$ and $\rho_2$ at 300 K for IrMn/YIG in Figs.~\ref{Fig2}(b) and~\ref{Fig3}(c), the ratio $G_i/G_r$ is evaluated to be 0.12, much larger than that (0.03) of Pt/YIG~\cite{Althammer2013}. After considering $G_r(\mathrm{IrMn})/G_r(\mathrm{Pt})=0.43$~\cite{Mendes2014}, one can see that $G_i$(IrMn) is larger than that of $G_i$(Pt) by a factor of 1.7, due to the low density of states near the Fermi level in IrMn~\cite{Sakuma2003,Xia2002}. The large $G_i$, indicating a large rotation of the reflected spin direction, provides more opportunities to manipulate the pure spin current~\cite{Xia2002}. \\
\begin{figure}[tb]
\begin{center}
\includegraphics[width=5cm]{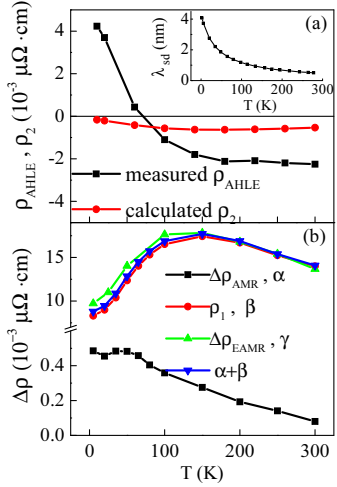}
\caption{For Pt/YIG, the $T$ dependence of measured $\rho_{AHLE}$ and calculated $\rho_2$ (a), and of $\Delta\rho_{AMR}$ (MPE AMR), $-\rho_1$ (SH AMR), and $\Delta\rho_{EAMR}$ (EAMR) measured by the angular dependence of $\rho_{xx}$ in the \emph{xz}, \emph{yz}, and \emph{xy} planes  under $H=10$ kOe (b). In the inset of (a), the $T$ dependence of the spin diffusion length of Pt is taken from Ref.~\onlinecite{Marmion2014}. For comparison, the sum of $\Delta\rho_{AMR}$ and $-\rho_1$ is also given. } \label{Fig4}
\end{center}
\end{figure}
\indent In summary, studies of the $T$ dependent AHLE and EAMR create a full understanding of the intricate MR properties in NM/insulating-FM heterostructures. The SMR and the MPE are both experimentally proved to be important in the mixed MR behavior. For IrMn/YIG, \textit{both} $\rho_{AHLE}$ and $\Delta\rho_{EAMR}$ change sign with $T$, due to the competition between the SMR and the MPE. For Pt/YIG, the sign change is observed only in $\rho_{AHLE}$ because the SH AHE is much weaker than the MPE AHE, and the SH AMR is much stronger than the MPE AMR. Moreover, the galvanomagnetic properties in NM/insulating-FM strongly depend on the magnetic attribute of the NM layer. Quite notably, it is the $T$ dependent AHLE and EAMR that make it easier to clearly map the physics behind the complex MR in Pt/YIG. The AHLE and the EAMR will also facilitate a better functionality and performance characterization of spintronic devices. \\
\indent This work was supported by the State Key Project of Fundamental Research Grant No. 2015CB921501, the National Science Foundation of China Grant Nos.11374227, 51331004, 51171129, and 51201114, Shanghai Science and Technology Committee Nos.0252nm004, 13XD1403700, and 13520722700. TEM specimen preparation and observation was performed at the Irvine Materials Research Institute at UC Irvine, using instrumentation funded in part by the National Science Foundation Center for Chemistry at the Space-Time Limit under grant no. CHE-0802913 \\

\end{document}


\title{Nature of magnetotransport in metal/insulating-ferromagnet heterostructures: Spin Hall magnetoresistance or magnetic proximity effect}

 \author{X. Zhou}
\affiliation{Shanghai Key Laboratory of Special Artificial Microstructure Materials and Technology and Pohl Institute of Solid State Physics and School of Physics Science and Engineering, Tongji University, Shanghai 200092, China}
\author{L. Ma}
\affiliation{Shanghai Key Laboratory of Special Artificial Microstructure Materials and Technology and Pohl Institute of Solid State Physics and School of Physics Science and Engineering, Tongji University, Shanghai 200092, China}
\author{Z. Shi}
\affiliation{Shanghai Key Laboratory of Special Artificial Microstructure Materials and Technology and Pohl Institute of Solid State
Physics and School of Physics Science and Engineering, Tongji University, Shanghai 200092, China}
\author{W. J. Fan}
\affiliation{Shanghai Key Laboratory of Special Artificial Microstructure Materials and Technology and Pohl Institute of Solid State
Physics and School of Physics Science and Engineering, Tongji University, Shanghai 200092, China}
\author{Jian-Guo Zheng}
\affiliation{Irvine Materials Research Institute, University of California, Irvine, CA 92697-2800, USA}
\author{R. F. L. Evans}
\affiliation{Department of Physics, University of York, York YO10 5DD, United Kingdom}
 \author{S. M. Zhou}
\affiliation{Shanghai Key Laboratory of Special Artificial Microstructure Materials and Technology and Pohl Institute of Solid State Physics and School of Physics Science and Engineering, Tongji University, Shanghai 200092, China}

\date{\today}
\vspace{5cm}

\maketitle
\section{Supplementary material}
\subsection{Fabrication and measurement details}
\indent  IrMn$_3$(=IrMn)/Y$_{3}$Fe$_{5}$O$_{12}$(=YIG) (20 nm) and Pt/YIG (20 nm) bilayers were fabricated on (111)-oriented, single crystalline Gd$_{3}$Ga$_{5}$O$_{12}$ (GGG) substrates. The base pressures of the PLD and sputtering systems were $1.0\times10^{-6}$ Pa. The YIG layer was epitaxially grown via pulsed laser deposition (PLD) from a stoichiometric polycrystalline target using a KrF excimer laser with the pulse energy of 285 mJ. The substrate temperature was 625 $^{\circ}$C during the deposition of the YIG layer. Then, the sample was annealed at the same temperature in an O$_2$ pressure of $6\times 10^4$ Pa for 4 hours. After the sample was cooled to the ambient temperature, it was transferred without the air exposure from the PLD chamber to the sputtering chamber through a load-lock chamber. Afterwards, the metallic layer was deposited at ambient temperature by DC magnetron sputtering, in order to avoid interfacial diffusion. The Ar pressure was 0.3 Pa during deposition of the metallic layer. The deposition rate of the metallic layer was about 0.1-0.2 nm/s.\\

\indent Structural properties and film thickness were characterized by X-ray diffraction (XRD) and X-ray reflection (XRR) using a D8 Discover X-ray diffractometer with Cu K$\alpha$ radiation (wavelength of about 1.54 {\AA}). The epitaxial growth of the YIG film was proved by pole figures with $\Phi$ and $\Psi$ scan at $2\theta$ fixed for the (008) reflection of the GGG substrate and YIG film. Transmission electronic microscopy (TEM) experiments were carried out in FEI/Philips CM-20 TEM with a LaB$_6$ filament operated at 200 kV at Irvine Materials Research Institute, University of California Irvine. Cross-sectional TEM specimens were prepared in a FEI Quanta 3D FEG dual-beam system with focused ion beam (FIB). A typical FIB procedure recommended by FEI Company was used to prepare the specimens. The thin film was well protected by two Pt layers deposited first by an electron beam and then by an ion beam. The final thinning step using a low energy (2 kV) ion beam is crucial to minimize an amorphous layer, a damaged layer caused by Ga-ion beam, on both sides of the TEM specimens. Magnetization hysteresis loops of the samples were measured using physical property measurement system (PPMS). The magnetization (134 emu/cm$^{3}$) of the YIG film is close to the theoretical value (131 emu/cm$^{3}$) and the coercivity is very small, 6 Oe. The films were patterned into normal Hall bar, and the transverse Hall resistivity ($\rho_{xy}$) and the longitudinal resistivity ($\rho_{xx}$) were measured by PPMS. \\
\subsection{XRD and XRR results}
\renewcommand{\thefigure}{S\arabic{figure}}
The XRR spectrum in Fig.~\ref{Figs1}(a) shows that YIG and IrMn layers are $20\pm 0.6$ and $5\pm0.5$ nm thick, respectively. The x-ray diffraction (XRD) spectrum in Fig.~\ref{Figs1}(b) shows that the GGG substrate and the YIG film are of (444) and (888) orientations. The pole figures in Fig.~\ref{Figs1}(c) and Fig.~\ref{Figs1}(d) confirm the epitaxial growth of the YIG film.
\begin{figure}[htbp]
\begin{center}
\resizebox*{5 cm}{!}{\includegraphics*{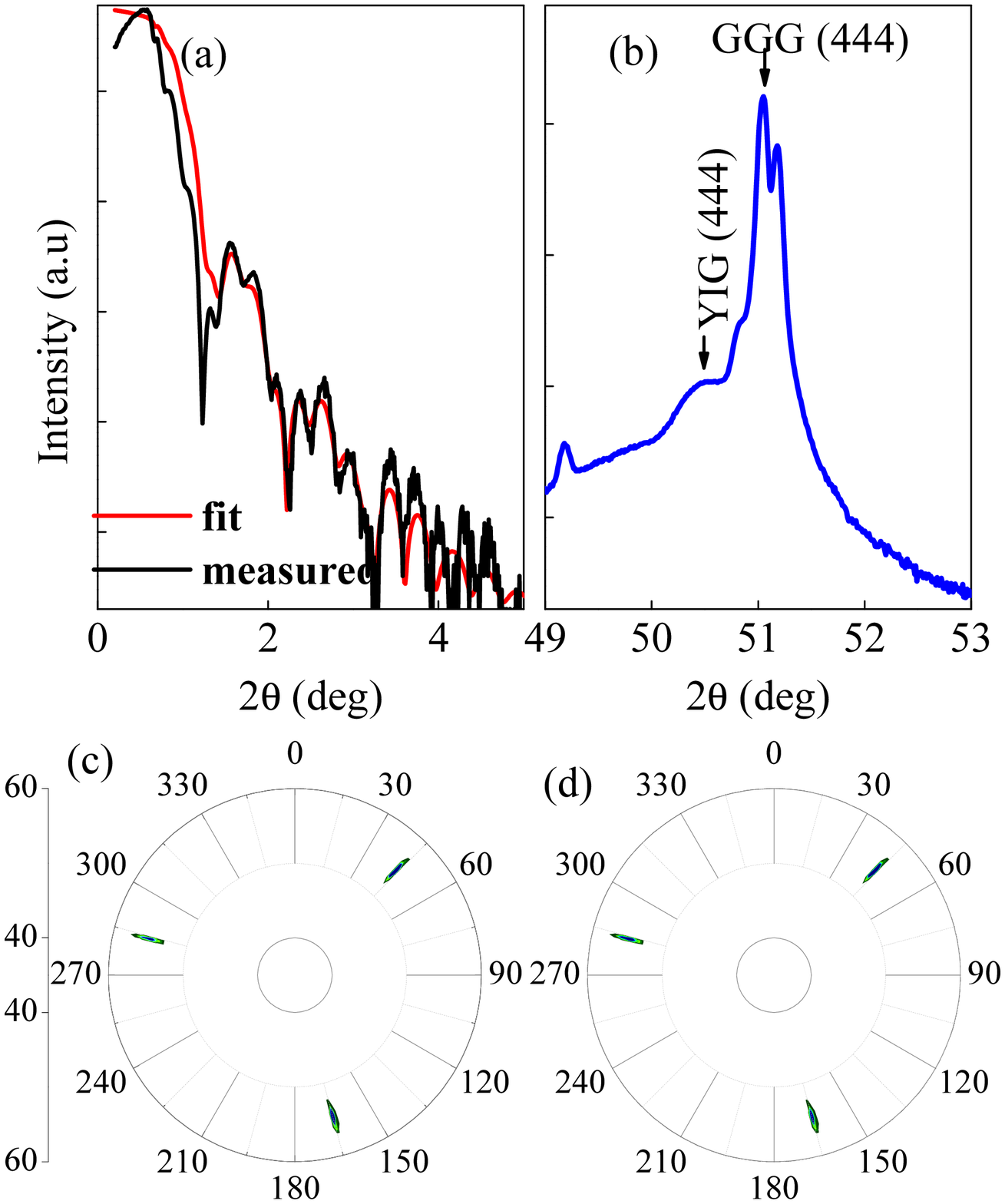}}
\caption{(a)Small angle x-ray reflection, (b)large angle x-ray diffraction for IrMn (5 nm)/YIG (20 nm) bilayer, $\Phi$ and $\Psi$ scan with fixed $2\theta$ for the (008) reflection of GGG substrate (c) and YIG film (d).}
\label{Figs1}
\end{center}
\end{figure}
\begin{figure}[htbp]
\begin{center}
\resizebox*{5 cm}{!}{\includegraphics*{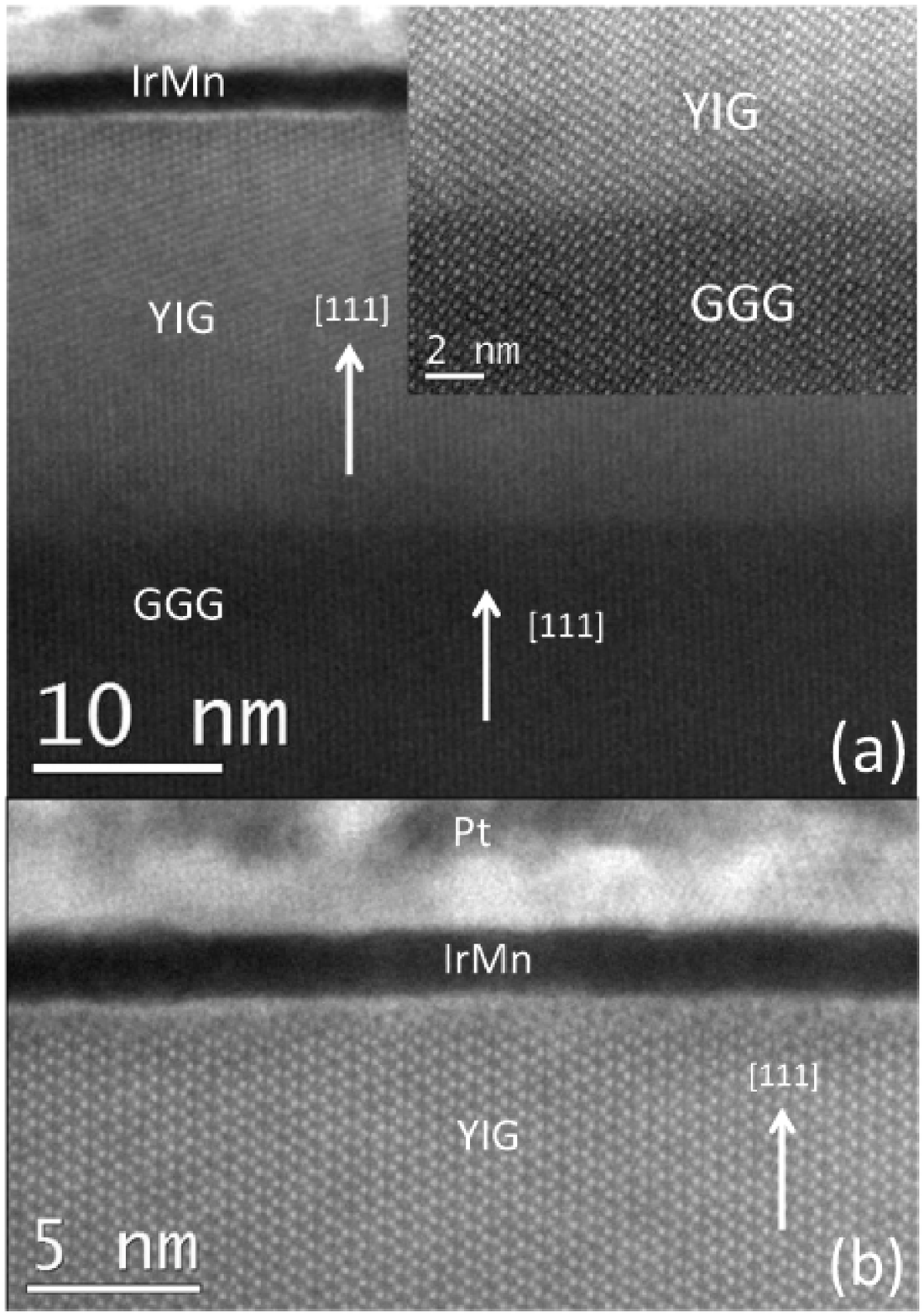}}
\caption{(a) Cross-sectional high resolution TEM (low magnification) image of IrMn/YIG bilayer on (111) GGG substrate, where the IrMn and YIG layers are 2.5 nm and 20 nm, respectively. The inset shows the atomic scale structure of the interface between YIG and GGG, indicating the epitaxial growth of YIG on GGG, (b) high resolution TEM (high magnification) image of the interface between IrMn and YIG.}
\label{Figs2}
\end{center}
\end{figure}
\subsection{TEM results}
Figure~\ref{Figs2}(a) shows a typical high resolution TEM image of the IrMn/YIG on the GGG substrate. The IrMn and YIG layers are characterized to be about 2.5 nm and 20 nm in thickness, respectively. The IrMn layer is polycrystalline, while the YIG layer is single crystal. The YIG is grown epitaxially on the GGG (111) substrate (inset of Fig.~\ref{Figs2}(a)). Detailed interface structure between IrMn and YIG is showed in Fig.~\ref{Figs2}(b). The fine Pt particles on top of the IrMn thin film form the protection layer which was deposited during TEM specimen preparation. The white area at the IrMn/YIG interface may be produced during the preparation of the TEM specimen because the milling rate at the interface is slightly higher than that in YIG. The interface between YIG and IrMn is flat and abrupt although the small YIG surface roughness at the interface is observed, where the root mean square surface roughness of the YIG layer is 0.35 nm.
\subsection{Temperature dependence of exchange bias}
\begin{figure}[htbp]
\begin{center}
\resizebox*{5 cm}{!}{\includegraphics*{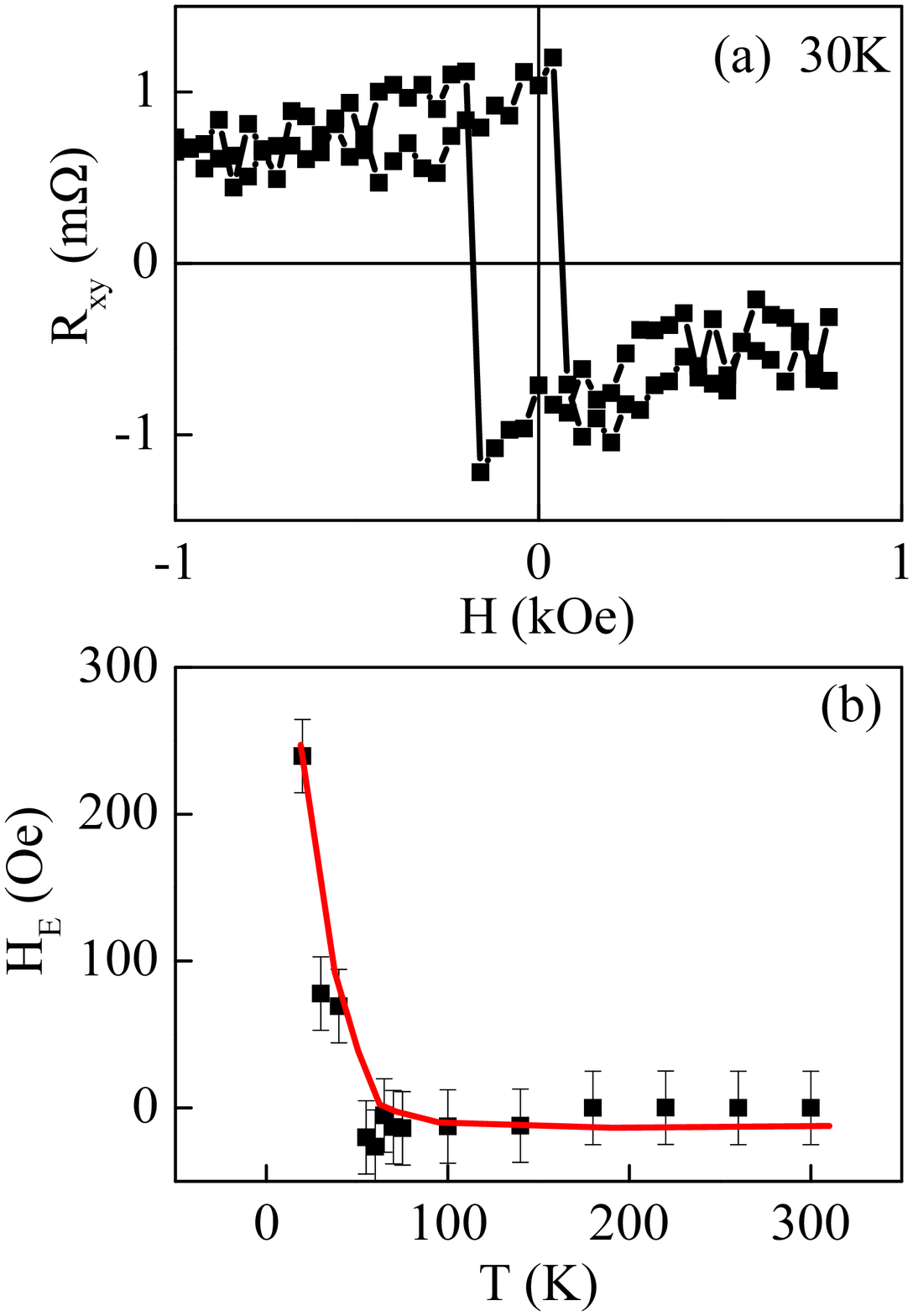}}
\caption{(a) In-plane hysteresis loop at 30 K and (b) temperature dependence of the exchange field for IrMn (2.5 nm)/YIG (20 nm) bilayers.}
\label{Figs3}
\end{center}
\end{figure}

\indent Figure~\ref{Figs3}(a) shows that for IrMn/YIG bilayers the hysteresis loop at 30 K is shifted away from zero magnetic field after a field cooling procedure under an in-plane magnetic field. The exchange field is 75 Oe at 30 K and decreases with increasing temperature $T$, and finally approaches zero near 70 K, as shown in Fig.~\ref{Figs3}(b). \\

\subsection{Atomistic simulations of the interfacial moment in $\gamma$-IrMn/YIG bilayers}
\begin{figure}[htbp]
\begin{center}
\resizebox*{8 cm}{!}{\includegraphics*{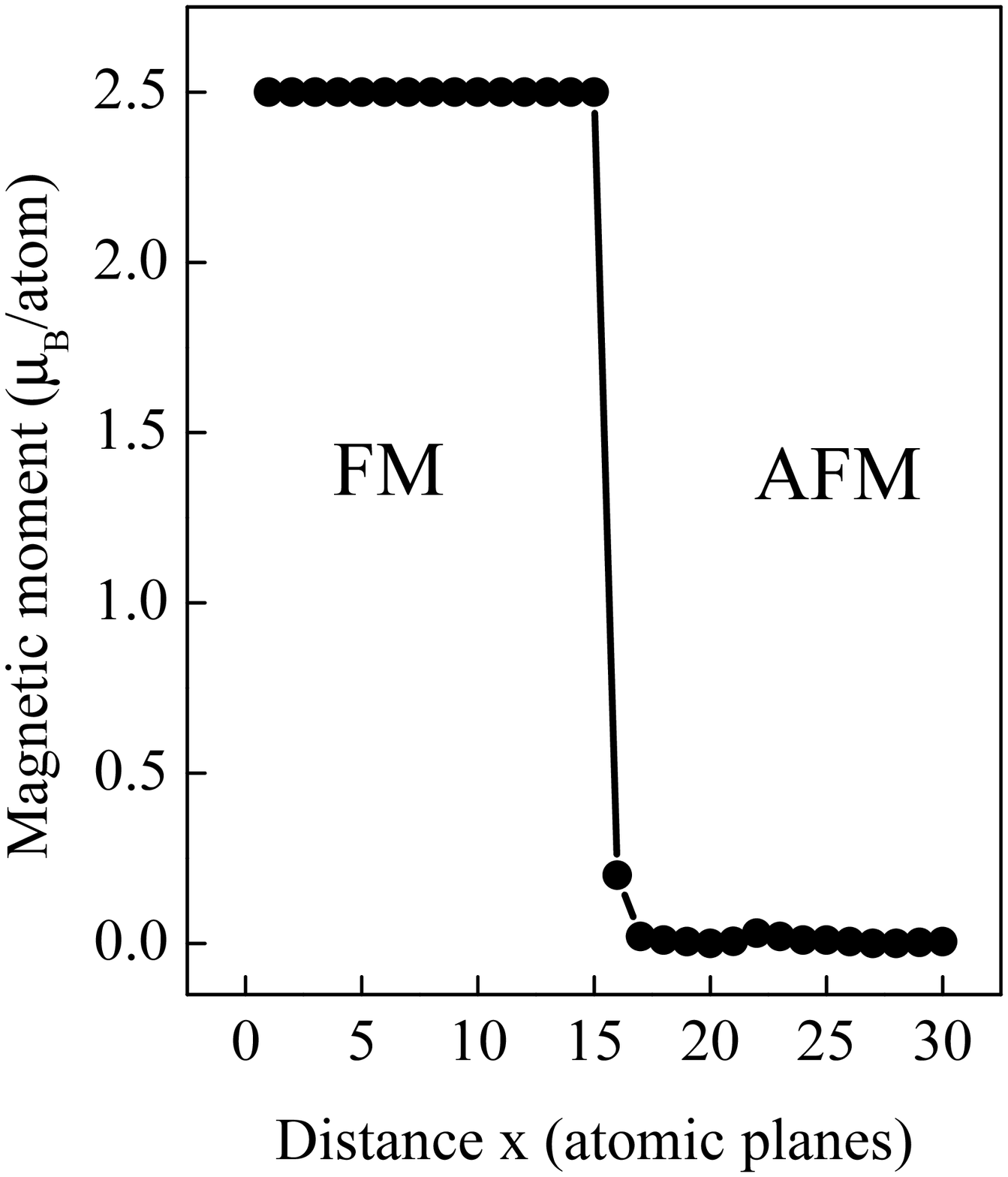}}
\caption{Atomistic calculation of the layer-wise total magnetic moment for a $\gamma$-IrMn/FM bilayer, showing an interfacial moment in the first atomic layer of IrMn.}
\label{Figs4}
\end{center}
\end{figure}

To confirm the presence of an interfacial spin moment in the IrMn layer, we perform atomistic spin dynamics simulations of an $\gamma$-IrMn$_3$/FM bilayer using the \textsc{vampire} software package\cite{Vampire}. The properties of the $\gamma$-IrMn$_3$ are modeled using a parameterized spin Hamiltonian which reproduces the 3Q magnetic ground state structure\cite{Jenkins2015}. A 20 nm $\times$ 20 nm $\times$ 3 nm thick IrMn layer is then coupled to a 3 nm thick generic ferromagnet with spin moment 2.5 $\mu_{\mathrm{B}}$. The magnetic ground state for the coupled system is determined by field cooling through the N\'eel temperature using the stochastic Landau-Lifshitz-Gilbert equation applied at the atomistic level\cite{Vampire}. The low temperature state is then analyzed to calculate the total magnetic moment in each atomic layer, as shown in Fig.~\ref{Figs4}.

The coupling of the two layers has a negligible effect on the FM magnetization, but induces a small magnetic moment in the first IrMn layer (layer 16) of around 0.2 $\mu_{\mathrm{B}}$ per spin on average. The second IrMn layers and beyond have no appreciable magnetic moment due to their antiferromagnetic structure, making the magnetic proximity effect in YIG/IrMn extremely short ranged. The atomistic simulations are applicable at the single gran level, but above the blocking temperature the local directions of the interfacial magnetic moment are randomized, leading to zero net magnetic proximity effect.

\subsection{Anomalous Hall conductivity versus IrMn layer thickness}
\indent At low $T$, the SH AHE almost disappears and only exists the anomalous Hall effect (AHE). Figure~\ref{Figs5} shows the dependence of the anomalous Hall conductivity (AHC) at 5 K on the IrMn layer thickness. The experimental data can be approximately fitted by the inverse proportion function of the IrMn layer thickness, indicating the interface nature of the AHC.   \\
\begin{figure}[htbp]
\begin{center}
\resizebox*{5 cm}{!}{\includegraphics*{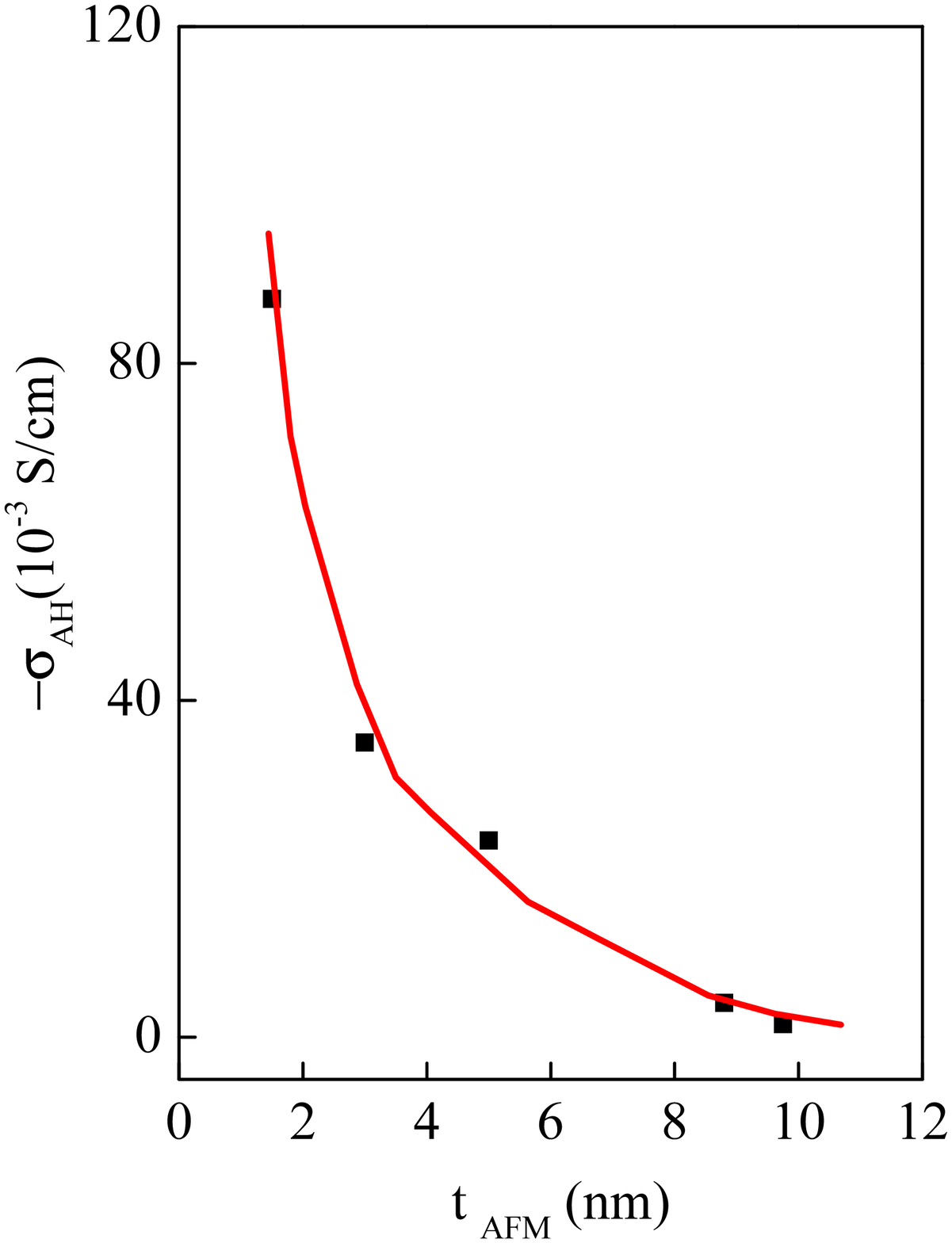}}
\caption{For IrMn/YIG (20 nm) bilayers, the AHC at 5 K versus the IrMn layer thickness.} \label{Figs5}
\label{Figs5}
\end{center}
\end{figure}